\documentclass[letterpaper]{article}
\usepackage{aaai}
\usepackage{times}
\usepackage{helvet}
\usepackage{courier}
\usepackage[hyphens]{url}

\usepackage{graphicx}
\usepackage{subcaption}
\usepackage{tabularx}
\usepackage{xcolor}
\usepackage{hyperref}
\usepackage{amsmath}
\usepackage{float}
\usepackage{dblfloatfix}
\usepackage{array}
\usepackage{booktabs}
\usepackage{varwidth}
\usepackage{balance}
\usepackage{makecell}
\graphicspath{ {./images/} }
\begin{document}

\title{Road to the White House: Analyzing the Relations Between\\ Mainstream and Social Media During the U.S. Presidential Primaries}

\author{Aaron Brookhouse$^*$\textsuperscript{1},
Tyler Derr$^*$\textsuperscript{2},
Hamid Karimi\thanks{Equal contribution and co-first authors.}\textsuperscript{1},
H. Russell Bernard \textsuperscript{3},
Jiliang Tang \textsuperscript{1}
\\
\textsuperscript{1}{Michigan State University},
\textsuperscript{2}{Vanderbilt University},
\textsuperscript{3}{Arizona State University}\\
\{brookho8,karimiha,tangjili\}@msu.edu,
tyler.derr@vanderbilt.edu, 
asuruss@asu.edu}
\maketitle
\begin{abstract}
\begin{quote}

Information is crucial to the function of a democratic society where well-informed citizens can make rational political decisions. While in the past political entities were primarily utilizing newspaper and later television to inform the public, with the rise of the Internet and online social media, the political arena has transformed into a more complex structure. Now, more than ever, people express themselves online while mainstream news agencies attempt to seize the power of the Internet to spread their agenda. To grasp the political coexistence of mainstream media and online social media, in this paper, we perform an analysis between these two sources of information in the context of the U.S. 2020 presidential election.  In particular, we collect data during the 2020 Democratic Party presidential primaries pertaining to the candidates and by analyzing this data, we highlight similarities and differences between these two main types of sources, detect the potential impact they have on each other, and understand how this impact relationship can change over time. To supplement these two main sources and to establish a baseline, we also include Google Trends search results and Polling results for each of the candidates that are being analyzed. 

\end{quote}
\end{abstract}

\section{Introduction}

The principal characteristic of a functional democratic society is having well-informed citizens that are expected to gain useful and accurate information so that they can make political decisions in a rational manner~\cite{nadeau2008election}. 
Then, a natural question that arises is where do citizens acquire information? In the past, the primary source of information was traditional unidirectional media such as newspapers or television where ``ordinary people" were mostly information consumers without access to a proper avenue to reflect their opinions. With the advent of the Internet in general and online social media in particular, however, we are now faced with a different and more complex scenario~\cite{dahlberg2007internet,weare2002internet}. On the one hand, mainstream news agencies have largely exploited the power of the Internet and fully expanded their news diffusion operations~\cite{newman2011mainstream}. On the other hand, thanks to readily accessible online social media platforms such as Twitter, a massive number of people express their political opinions online~\cite{jungherr2014twitter,tumasjan2010predicting}. 
In addition to the sheer high volume of information produced online, these two main sources of information (i.e., mainstream news outlets and online social media) are interacting and influencing each other. Hence, a thorough analysis of the symbiosis of online social media and mainstream news is highly desired and recent efforts have focused on this in other domains~\cite{jang2019social}. 
To perform this analysis, we focus on a crucial political event, namely the 2020 U.S. presidential primaries/election. In the following, we further explain the relationship between different news sources and their role in elections.

Each of the sources of information can affect people's decisions in an election and they can also influence each other. It suggests that the impact of a news source is beyond its own readers. Online social media is shown to make individuals' political opinions farther apart from each other, yet it is also linked to individuals with more exposure to beliefs of people on the other side of the political spectrum \cite{flaxman2016filter}. Moreover, ideally, the news should only serve as a source of information that accurately reports the events of the world. However, in reality, it influences what topics people are interested in, how the topics are thought about, and perhaps following some sort of agenda~\cite{parenti1993inventing}, which is an area of research existing long before modern social media~\cite{berkowitz1987tv,roberts1994agenda,berkowitz1992sets} and commonly known as agenda-setting in communications research.
It is important to be aware of this when consuming any sort of media. In fact, one concern is that it has been shown that fake news on Twitter can travel up to six times faster than real stories, and furthermore it is humans, not bots, that are the primary spreaders of this misinformation~\cite{vosoughi2018spread}. On the other hand, social media campaigns have had impressive results, shaping the public discussions among many topics, and have had a large influence on public opinions as well as the news cycle~\cite{cogburn2011networked,metzgar2009social,kushin2010did}. It has also been shown through polls that people who align with a particular party are more prone to certain misbeliefs about recent events, and conservatives and liberals get their news from different sources~\cite{polarizationBiasModel}. In fact, most online news traffic is from people visiting the homepage of their favorite newspaper's sites~\cite{flaxman2016filter}.

In this paper, we will explore the relations between different sources of information under the context of the 2020 U.S. presidential primaries/election from various perspectives. These sources include mainstream news, Twitter, Google Trends, and polls. First, we perform an initial analysis of the data distributions and how their time series are correlated. Second, we determine how each candidate is portrayed in conservative, liberal, and neutral news sources. Moreover, we determine the public's sentiment toward each candidate using Twitter data. Third, we investigate the co-correlations among the candidates in both news and Twitter data to better understand how the candidates are related in the two forms of media. In particular, on a given platform, how correlated the mention frequency is for two candidates. This helps to also understand how candidates may have been perceived differently on different platforms. 
Fourth, we perform a cross-media influence where we identify how one source influences the other. Moreover, since the interaction between sources is crucial, we further investigate the influence between Twitter and mainstream media from a casual perspective where we attempt to predict the trend in one source using the other and vice versa.
Fifth, we strengthen our cross-media influence by using the Granger Causality Test to quantify this relationship and test if one source is adapting or reacting to changes in the other source.  
Sixth, we analyze how topics of interest are different across platforms, and candidates. This indicates which topics or issues a candidate is most perceived to be related to on a given data platform coming from both the aspects of the content they are creating and what topics the public is mentioning when discussing the candidate. Finally, we perform a toxicity analysis where for each candidate we determine the degree of toxicity of their mention on Twitter. Our main contributions of this work are summarized as below:

\begin{itemize}
    \item We present the first comprehensive analysis of the two major information sources that are being used to impact the 2020 U.S. presidential primaries/election, by focusing on Twitter and several mainstream news agencies.
    \item Each of our described analyses is presented from two perspectives with two goals. They provide direct insight into the candidates themselves as they relate to the news, public opinion, and the upcoming election. We also aim to highlight the difference between news and public opinion from a candidate-agnostic standpoint.
    \item We introduce approaches that nuanced ideas such as bias, the topic mismatch between sources, influence, and toxicity that can be quantified.
\end{itemize}

In the next section of the paper, we will further explain exactly how the data collected and provide an overview of the raw data. Then, for each of the analyses described above, we will further explain and demonstrate the processes involved and present the associated results. Finally, we present a summarized conclusion of our findings and discuss future directions that emerge from this investigation.

\begin{table*}[t]\small
    \setlength{\extrarowheight}{1.25pt}
    \setlength\tabcolsep{3.5pt}
    
    \centering
    \caption{Twitter and news data statistics}
    \begin{tabular}{|c||c|c|c|c|c|c|c|c|c|c|c|c|}
    \hline 
         \textbf{Candidate} &\makecell{ Elizabeth\\ Warren} & \makecell{ Bernie\\ Sanders} & \makecell{Pete\\ Buttigieg} & \makecell{Kamala\\ Harris} & \makecell{Beto\\ O'Rourke} & \makecell{Cory\\ Booker} & \makecell{Andrew\\ Yang} & \makecell{Amy\\ Klobuchar} & \makecell{Tom\\ Steyer} & \makecell{Donald\\ Trump} & \makecell{Joe\\ Biden} \\ \hline
         \textbf{\makecell{Twitter \\ Average\\ \#TPD}}& 53,286    &  127,213       & 31,938       &  16,808       &  3,086      & 4,526     & 37,585       &  12,020 & 4,194  &  160,905 &  138,923 \\ \hline
         \textbf{\makecell{News \\ Average \\ \#APD}} &  7.66     &  18.70     &  6.18      &  2.23    &  0 .21     & 1.20     & 1.83      & 2.42      &  0 .94       &  141.23    & 25.07  \\ \hline
         \textbf{\makecell{Ratio}} & 6,956   &  6,803     & 5,168      & 7,537      & 14,695       & 3,772  & 20,538   & 4,967    & 4,462      & 11,393    & 5,541        \\ \hline
    \end{tabular}
    \label{tab:twitternewsdata}
\end{table*}

  \begin{figure*}[t]
    \vskip -1ex
    \hspace{-2ex}
    \includegraphics[width=1.04\textwidth]{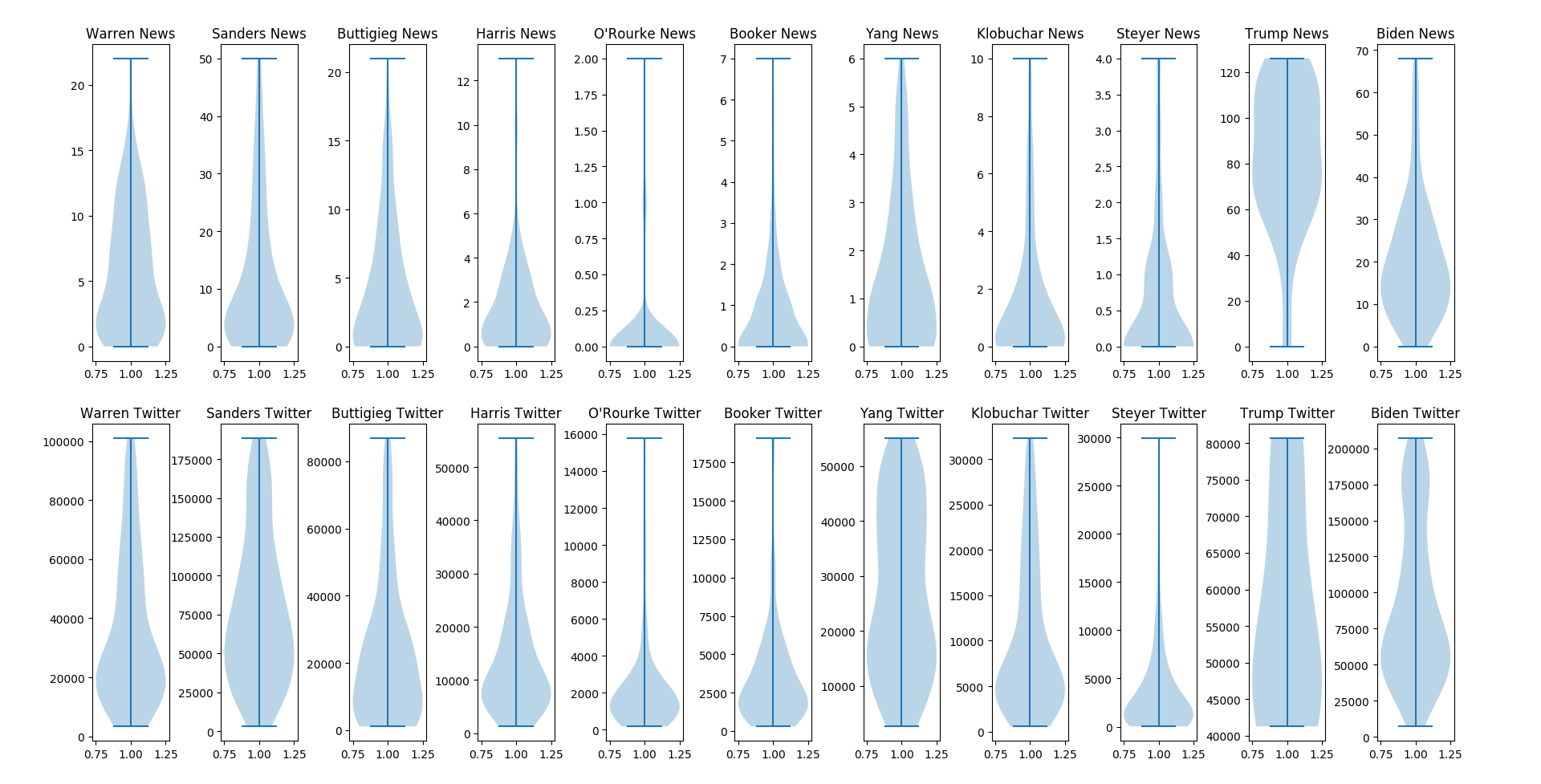}
    \captionof{figure}{Visualizing the distribution of the documents related to each candidate per day.}
    \label{fig:violin}
  \end{figure*}

\section{Data}

There are four main sources of data that we are analyzing: news, Twitter, Google Trends, and polls. For the news, data was collected from Currents API\footnote{https://currentsapi.services/}, which provides information about articles such as title, URL, time published, and an excerpt of the article. To obtain the full texts of the articles, the Python Newspaper API was used\footnote{ https://newspaper.readthedocs.io/}. The Twitter data was collected using the Tweepy Python Streaming API\footnote{https://www.tweepy.org/}. A program was written to download tweets that mentioned any of the Presidential candidates in real-time. For Donald Trump, only approximately 10\% of tweets were saved given the fact that his mentions were significantly more than the Democratic primary candidates. Specifically, since there are over 160,000 tweets collected per day, this should be sufficiently representative for our study. We collected Google Trends data directly through their online tool\footnote{https://trends.google.com/trends/}. The poll data was taken from RealClearPolitics\footnote{https://www.realclearpolitics.com/epolls/2020/president/us/ 2020\_democratic\_presidential\_nomination-6730.html} where we collected their averaged poll data to avoid the bias of a single polling source.

We further process the Twitter and news data by extracting the following information: 1) the number of documents per day, 2) average sentiment of documents each day, 3) a point series of document publication times, and 4)  political topics of the documents.
In Table~\ref{tab:twitternewsdata}, we summarize the average amount of documents of each candidate per day, i.e., tweets per day (TPD) and articles per day(APD), for Twitter and news, respectively. A clear observation from Table~\ref{tab:twitternewsdata} is that both Twitter and the news are quite variable in the number of documents appearing on average per candidate. 
Upon further investigation, we observed that the documents per day distributions were not following a normal distribution, but in many cases had a long tail distribution. Hence, to better understand the Twitter and news document frequency data, we visualize the data distributions in Figure \ref{fig:violin}. We note that different candidates do indeed follow different distributions of coverage not only between each other but also between sources.

\begin{table*}[t]
 \vskip -1ex
    \small

    \setlength{\extrarowheight}{3pt}
    \setlength\tabcolsep{1.5pt}

\centering
\caption{Pearson Correlations between data sources for each candidate.
\label{tab:differentMediumCorrelation}
}
\begin{tabular}{|l|c|c|c|c|c|c|c|c|c|c|c|c|} 
\hline
Pairwise Relations
&
\begin{tabular}{@{}c@{}}Elizabeth \\Warren \end{tabular}
&
\begin{tabular}{@{}c@{}}Bernie \\Sanders\end{tabular}
&
\begin{tabular}{@{}c@{}}Pete \\Buttigieg \end{tabular}
&
\begin{tabular}{@{}c@{}}Kamala \\Harris\end{tabular}
&
\begin{tabular}{@{}c@{}}Beto \\O'Rourke \end{tabular}
&
\begin{tabular}{@{}c@{}}Cory \\Booker\end{tabular}
&
\begin{tabular}{@{}c@{}}Andrew \\Yang \end{tabular}
&
\begin{tabular}{@{}c@{}}Amy \\Klobuchar\end{tabular}
&
\begin{tabular}{@{}c@{}}Tom \\Steyer\end{tabular}
&
\begin{tabular}{@{}c@{}}Donald \\Trump\end{tabular}
&
\begin{tabular}{@{}c@{}}Joe \\Biden \end{tabular}
&
\begin{tabular}{@{}c@{}}Avg. \end{tabular}\\ \hline

News - Twitter & .78 & .84 & .80 & .79 & .57 & .74 & .50 & .62 & .66 & .17 & .74 & .74 \\ \hline
News - Google Trends & .24 & .64 & .37 & -.01 & .47 & -.04 & .06 & .49 & .33 & -.02 & .30 & .26 \\ \hline
Twitter - Google Trends & .59 & .70 & .45 & -.07 & .21 & .01 & -.01 & .44 & .27 & .13 & .12 & .26  \\
\hline
\end{tabular}
\end{table*}

\begin{figure*}[!t]
\vskip -1ex
 \centering
  \includegraphics[width=.97\textwidth]{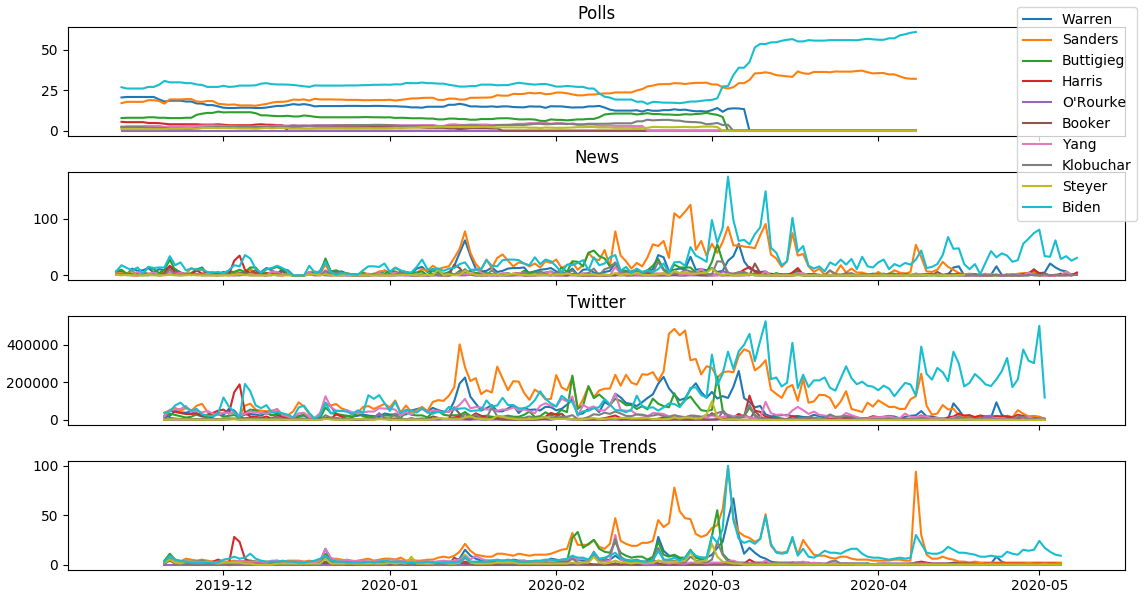}
  \vskip -1ex
 \caption{Candidates' poll, news, Twitter, and Google Trends data spanning the major points of the 2020 Democratic primaries.}
 \label{fig:GoogleTrends}
 \end{figure*}

\subsection{Time Series Analysis Between Data Sources} 
In Figure~\ref{fig:violin}, we had presented the distributions of the Twitter and news data to observe the difference in trends per candidate in the two mediums. However, given that this data is inherently time series, in Figure~\ref{fig:GoogleTrends} we also plot the candidates' data over time from the four sources we have collected, namely polls, news, Twitter, and Google Trends. 

The first observation we make from Figure~\ref{fig:GoogleTrends} is that News, Google Trends, and Twitter data follow a similar trend. For example, this is even more obvious when looking at the obvious spikes of Bernie Sanders, Joe Biden, and Elizabeth Warren. We observe that these are positioned around major points during the primaries, such as the seventh Democratic Party debate in the middle of January, Bernie Sander's early lead in February, Elizabeth Warren ending her campaign after unsatisfactory results in Super Tuesday along with Joe Biden taking the lead after receiving the endorsement of many candidates (such as Pete Buttigieg) who ended their campaigns before Super Tuesday, and finally Bernie Sander's ending his campaign in April of 2020. We elected to use 
Google trends since they directly measure the frequency of search terms and is a reliable gauge of the public's interest. Given that Google Trends shares similar trends with Twitter, the frequency of Tweets is justifiable to use as a representative measurement of the public's interest. Others have also demonstrated that Google Trends and Twitter are good indicators of public interest~\cite{trendTwitterRealOpinion}.  
In an effort to more precisely quantify and measure the level of similarity between the news, Twitter, and Google Trends, in Table \ref{tab:twitternewsdata} we present the pairwise Pearson correlation coefficients between these different sources at the candidate level and provide an averaged column across the candidates. The first thing to note is that on average News and Twitter are the two most correlated of the three data sources. The next interesting is that while some candidates such as Amy Klobuchar and Bernie Sanders have relatively stable and higher correlations among all pairs, others such as Kamala Harris and Joe Biden do not. More specifically, we observe that both Kamala Harris and Joe Biden had a high correlation between news and Twitter, but the near-zero correlation between Twitter and Google Trends (and very low for News and Google Trends as well). The highest correlation among all the three pairwise relations was Bernie Sanders. Lastly, we note that Donald Trump was significantly below average across all three pairwise data source correlations.

 \begin{figure*}[t]
 \centering
  \includegraphics[width=0.85\textwidth]{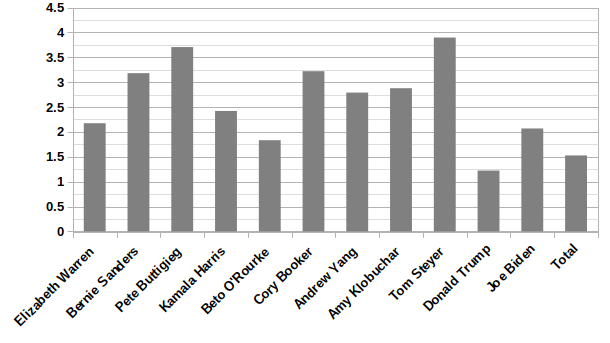}
  \vskip -2ex
 \caption{Overall news article sentiment taking the ratio of positive to negative articles per candidate.}
 \label{fig:newspostoneg}

 \end{figure*}

\section{Sentiment Analysis Per Candidate From Biased News Sources}

It is well known that different news sources have different biases, with some having a conservative viewpoint, others having a liberal viewpoint, and some being somewhere closer to the middle. Although it would seem intuitive how these different news sources might portray certain candidates (based on their political affiliation and personal ideologies), it is hard to make objective observations about these differences between these sources in order to understand the media better. Thus, a specific quantifiable measurement is desired to guide our understanding of these differences. In this work, we utilize sentiment analysis~\cite{liu2012survey}.

More specifically, to quantify some of these differences, we will look at differences in sentiment in news articles that include the different candidates for the 2020 Democratic nomination along with the current U.S. President - Donald Trump. To measure sentiment, we elected to use the popular method VADER\footnote{https://pypi.org/project/vaderSentiment/} which is a lexicon and rule-based sentiment model~\cite{vader}.

 \begin{figure}[t]
 \centering
  \includegraphics[width=0.49\textwidth]{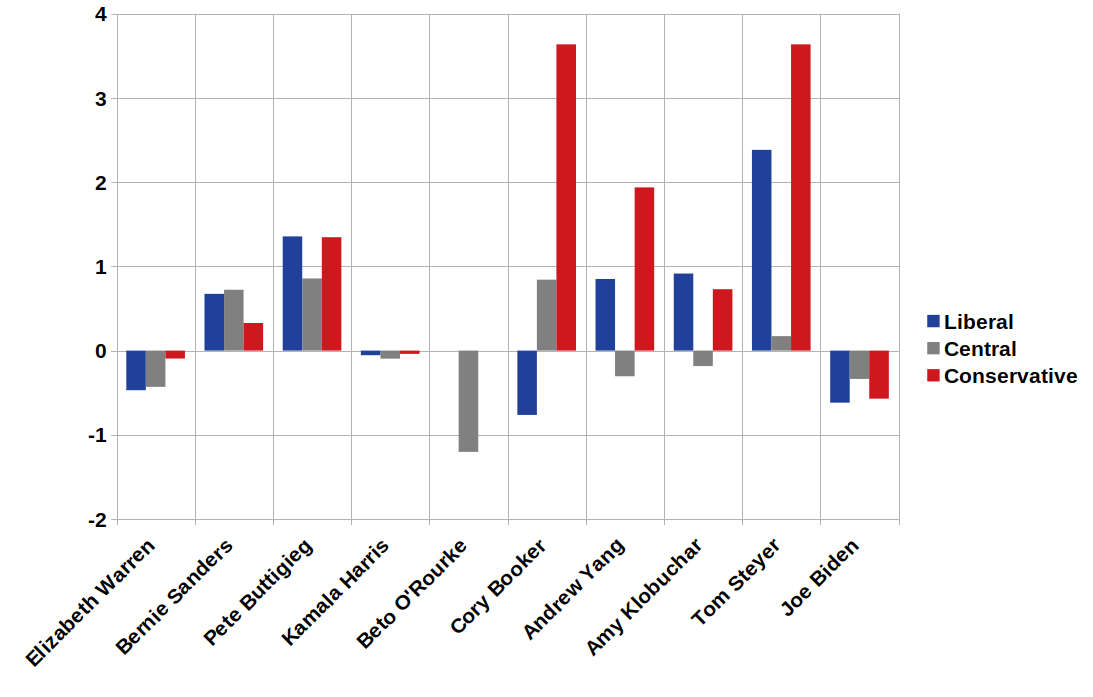}
  \vskip -2ex
 \caption{Ratio of positive to negative articles per candidate, with average shifted to zero.}
 \label{fig:newsDelta}
 \vskip -2ex
 \end{figure}

 Figure~\ref{fig:newspostoneg} presents our initial findings showing the ratio of positive to negative articles after having classified each of the candidate's articles across all news sources. We can observe that overall on average the mainstream news is providing a positive view for each of the candidates and current President (although we note that Donald Trump had the lowest ratio). We also note that the four highest each surpassing the ratio of 3.0 were Tom Steyer, Pete Buttigieg, Cory Booker, and Bernie Sanders.

 Next, we utilize the categorization of the news sources to present a second viewpoint as compared to an overall aggregate, we only aggregate together news articles within the same category of liberal, central, and conservative (as previously mentioned). We again perform an analysis of the ratio of positive to negative articles, but this time taking into consideration the category of the news source and shifting the axis to have zero be the average, in Figure~\ref{fig:newsDelta}.  
 Interestingly these results also do not completely follow party lines, since when we analyzed Donald Trump separately his conservative media has a lower ratio of 1.17 than liberal media having 1.36. When running the daily average sentiment from liberal news and conservative news in the two-sample t-test, we observe a t-value of -2.9, and a p-value of .0039, which indeed suggests a significant difference. This means that even though the conservative news sources officially support him, the articles they write including him are overall more negative than the other news sources. There are also many Democratic candidates who have higher conservative sentiment scores than liberal sentiment scores. We also note that some of the candidates such as Cory Booker and Tom Steyer have a very high ratio among the conservative news sources, but could potentially also be noisy, since they had very few articles (i.e., on average roughly one news article per day in total) and were candidates that seemed less likely to win from the beginning to the end of their campaigns. This may indicate that the more moderate candidates and less well-known candidates are preferred by the conservative media. This theory appears to also apply in reverse to the more popular and well-known candidates. Joe Biden and Bernie Sanders, the two most popular and well known Democratic candidates have lower conservative sentiment scores than they have from liberal and central sources. 
 
  \begin{figure}[t]
     \centering
     \includegraphics[width=.48\textwidth]{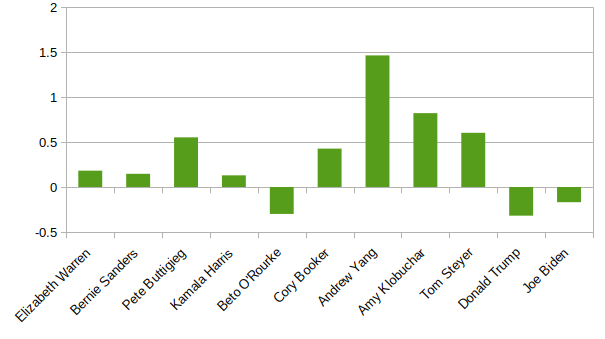}
     \vskip -1ex
     \caption{Ratio of positive to negative Tweets per candidate, with average shifted to zero.} 
     \label{fig:twitterpostoneg}
     \vskip -1ex
 \end{figure}
 
 It is also interesting to note, that in many instances, and also overall, the liberal-leaning news sources have a slightly higher average sentiment. Second is conservatively leaning news, with more central news sources having the lowest average sentiment. When the combination of two of these three types of news sources (Liberal, Neutral, and Conservative) are assessed through the two source t-tests, the pair Liberal and Conservative gets a p score of .0001 and the pair Neutral and Conservative has a p score of .0049. Thus, both pairs have an average sentiment that is significantly different from the other. On the other hand, the pair Liberal and Neutral though have a p score of .28, indicating that the difference in average sentiment between them is not significant. 

For completeness, we also calculate the ratio of positive to negative tweets per candidate in Figure~\ref{fig:twitterpostoneg}. We notice less variation around the mean in Twitter as compared to the mainstream news articles (shown in Figure~\ref{fig:newsDelta}). Another thing to note is that although the campaigns for Andrew Yang and Amy Klobuchar were not successful enough to allow them to stay in the race, they actually had the highest ratio of positive to negative Tweets (while also not even in the bottom three for average Tweets per day).

\begin{figure*}
     \centering
     \begin{subfigure}{0.49\textwidth}
         \centering
         \includegraphics[width=1.0\linewidth]{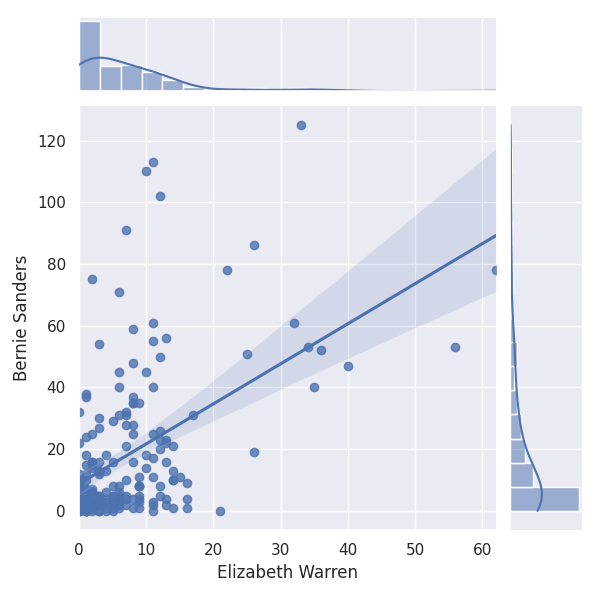}
         \caption{News (r=.50)}
         \label{fig:NewsCoCorrelationExample}
     \end{subfigure}
     \hfill
     \begin{subfigure}{0.49\textwidth}
         \centering
         \includegraphics[width=1.0\linewidth]{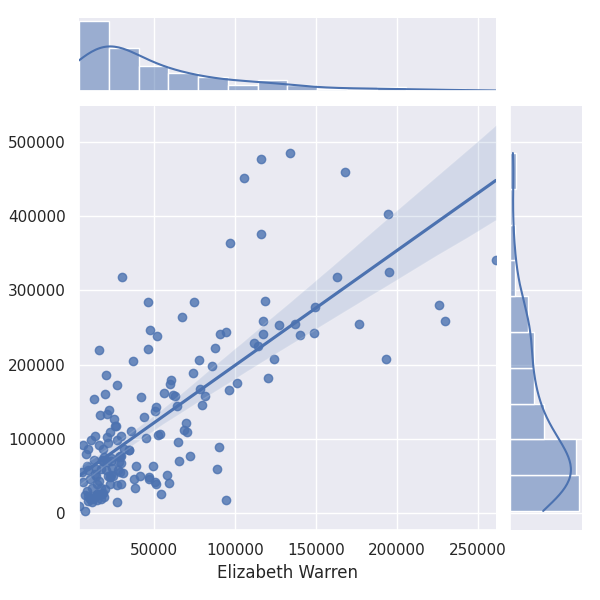}
         \caption{Twitter (r=.75)}
         \label{fig:TwitterCoCorrelationExample}
     \end{subfigure}
 \caption{Visualizing the frequency co-correlation between Bernie Sanders and Elizabeth Warren.}
 \label{fig:extra}
\end{figure*}


  \begin{figure*}[t]
 \centering
  \includegraphics[scale=0.25]{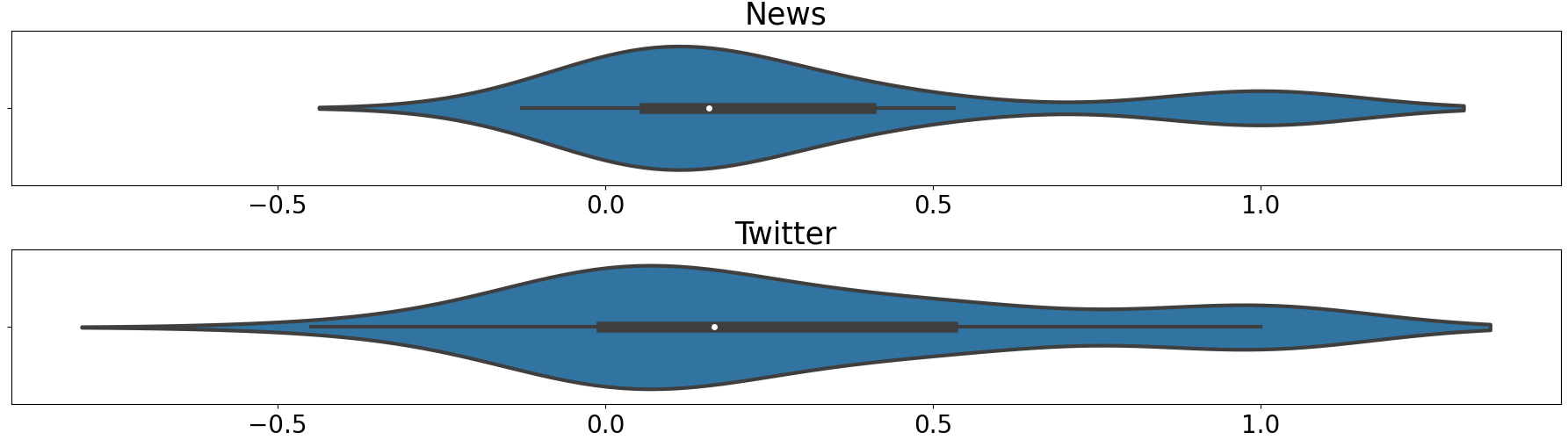} 
 \caption{Distribution of co-correlations among all candidate pairs in news and Twitter.}

 \label{fig:cocorrelationViolin}
 \end{figure*}

\section{Candidate Co-Correlations}

One type of influence that a media can exert on an observer is to group candidates together. This can impact how a viewer perceives the candidates. For example, if two candidates are often mentioned together as opposing each other, 
could a positive thought about one lead to a negative thought about the other? Or in the opposite case, two candidates often being mentioned together agreeing may cause a viewer to have a thought about one that they then apply to the other automatically. This type of grouping of ideas decreases objectivity, as what one candidate does or says should not automatically impact perceptions of another candidate. To remain most objective, media sources should not consistently group candidates together. They should rely on information that pertains to each candidate separately without impressing upon the media viewer that two ideas are related. Also, it could potentially be a strategy to construct candidates that are designed to appeal to a certain smaller demographic and then by design have them later endorse (or strongly oppose) another candidate.  

We decided to measure this level of grouping of candidates by a media platform by their co-correlations. Candidates were divided up into all possible pair combinations. For the candidate pair (X, Y) two figures were made, one for Twitter and one for news. For each day, a data point is plotted on the figure. The x-axis denotes the number of articles/tweets published about candidate X, and the y-axis denotes the same for candidate Y. The Pearson correlation is calculated for each pair of these co-correlations.

For a case study, we chose the pair Elizabeth Warren and Bernie Sanders to visualize the difference in variance between Twitter and news because they show a strong co-correlation in both Twitter and news as seen in Figures \ref{fig:TwitterCoCorrelationExample} and \ref{fig:NewsCoCorrelationExample} (having correlation values of 0.75 and 0.50, respectively). Some candidate pairs have stronger correlations than others -- for some pairs, the co-correlation is stronger with the news data than with the Twitter, while some instances are the other way around. This relationship is shown in Figure~\ref{fig:cocorrelationViolin} where the distributions of co-correlations are plotted. These averaged co-correlations are low with 0.1662 in news and 0.1309 in Twitter. This is because many candidates had very low co-correlation, while some had much higher (such as Elizabeth Warren and Bernie Sanders). While the news average co-correlation is higher by about 25\%, a two-input t-test measures a p-value of .75, so there is not a significant difference in candidate grouping tendencies between news and Twitter sources overall (although we can visualize some differences in the distributions in Figure~\ref{fig:cocorrelationViolin}). Overall, this co-correlation can show which candidates are portrayed/perceived as related to each other. Here, we demonstrate that there is not a tendency for news or Twitter to be more likely to group candidates together than the other. 

\section{Cross-Media Influence}
In today's world, every day we have different forms of media influencing each other. This means that even if you are not directly observing a certain platform of media, you are still likely susceptible to its influence on certain topics. This suggests that people who don't use social media are still indirectly affected by what people say about the candidates. Also, people who don't read the mainstream news online are still indirectly impacted by what it has to say about different candidates. Furthermore, it provides insights on whether it is the public opinion (such as that found in social media) influencing mainstream news, mainstream news influencing public opinion, or bidirectional and to what extent are they influencing each other. To do this, we quantify the influence of cross-media platforms.

\begin{figure}[t]
 \centering
  \includegraphics[width=0.5\textwidth]{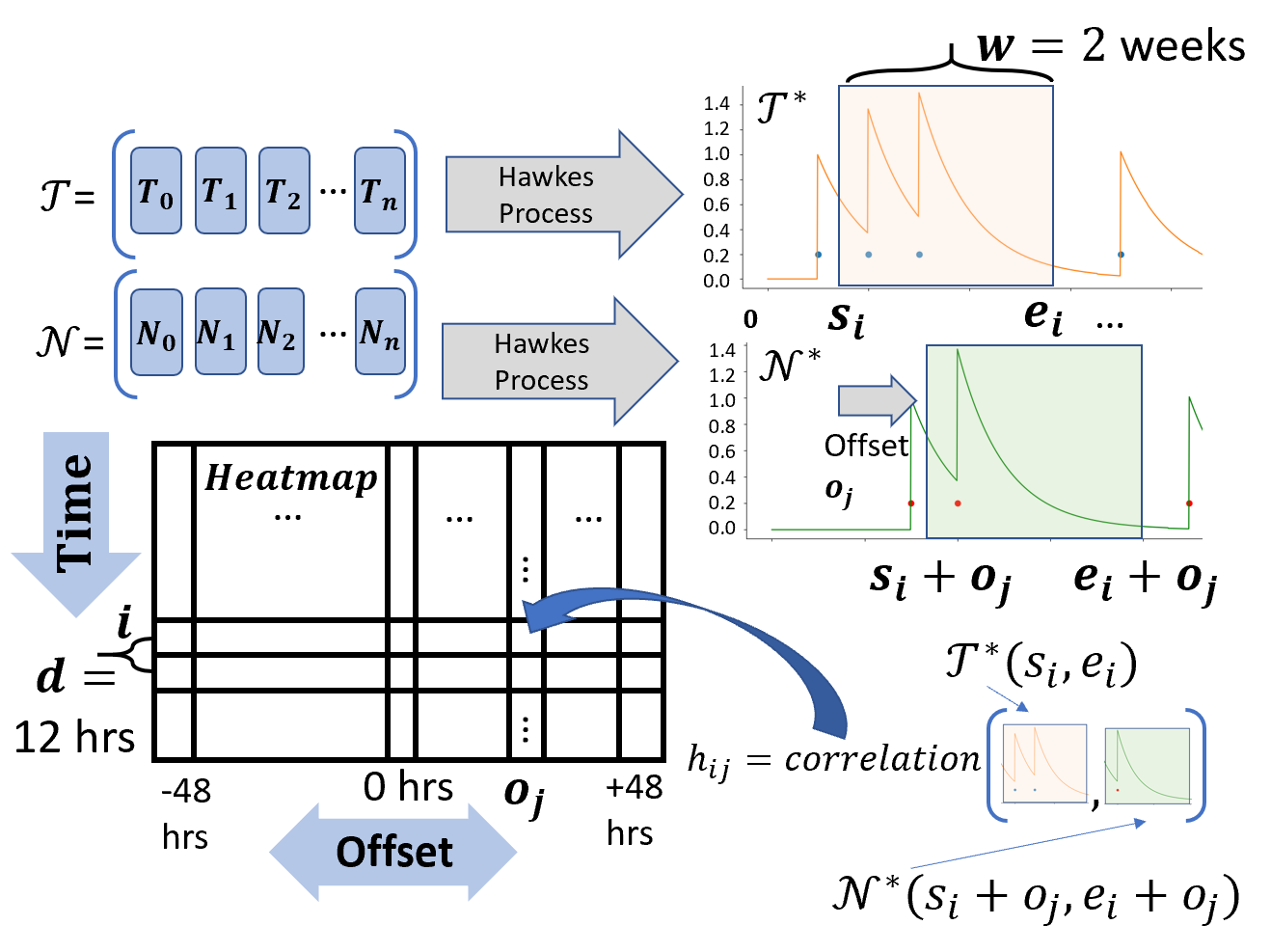}
  \vskip -1ex
 \caption{Explanation of the correlation heat maps.}
 \label{fig:heatmapExplanation}
 \vskip -1ex
 \end{figure}

More specifically, one way to determine which of two media sources is influencing the other more is to shift the data such that the time $t$ on one dataset will align with time $t-o$ of the other dataset, where $o$ is a defined offset in time. To gain insights into the influence, we can vary the offset and calculate the Pearson correlation coefficient. 
In this way, the offset points at which the correlation coefficient is greatest shows where one source of media is influencing the other.
If the coefficient, for example, is highest when data $x$ is leading data $y$, then it can be implied that data $x$ is influencing data $y$. To analyze these relations, we create heatmaps showing these correlations to better understand the effect over time and based on multiple offsets. In Figure~\ref{fig:heatmapExplanation}  we visualize this process and will next introduce it more formally while pairing the explanation with this figure.

First, we perform this investigation by binning the Twitter and news data into 5-minute bins denoted as $\mathcal{T}=[T_0, T_1, \cdots T_n]$ and $\mathcal{N}=[N_0, N_1, \cdots N_n]$, respectively, where we assume to have $n+1$ 5-minute bins in our data. Note that in our data the initial starting time is November 11, 2019, and the ending time is May 14, 2020. To better understand the potential influence between Twitter and news over time, we can localize two-week windows, i.e., $w=2$ weeks, and obtain the correlation between them locally (as compared to the entire length of time) i.e., each row of the matrix in Figure~\ref{fig:heatmapExplanation}. Then, we can run this window across our entire data over time obtaining multiple views of the correlation (i.e., going from the top to bottom row of the matrix in Figure~\ref{fig:heatmapExplanation} with offset as 0 hours). While sliding the window of size $w$ we shift in intervals of $d=12$ hours. 
However, as mentioned before, it is also of interest to obtain the correlation when leading one data before the other to uncover potential influence that might exist in the data. Hence, we utilize a set of offsets $\mathcal{O} = \{ -48 hrs, -47 hrs, \cdots, 48 hrs\}$ that range from -48 hours to 48 hours in one-hour intervals, which are added to one of the data sources (e.g., News) while keeping the other one fixed (e.g., Twitter) (i.e., adding an offset is moving horizontally away from the matrix central column in Figure~\ref{fig:heatmapExplanation}. 

However, when attempting to zoom in to such a fine-grained view, it is possible that many hour intervals will not have news articles being created for some of the lesser discussed candidates. Thus, to alleviate this, we harness the exponential decay of the Hawkes Process (which is commonly applied to Twitter and news point series data~\cite{rizoiu2017tutorial}). This allows us to convert our discrete frequency series of when tweets/articles are being created over time to a single continuous line through the use of exponential decay. In this case, the publishing times of each tweet/article are fed into the Hawkes Process, and a smooth line with many data points is output (as seen in Figure~\ref{fig:heatmapExplanation}). At times, the news articles are published somewhat sparsely, especially compared to Tweets about the different candidates. The Hawkes process helps to smooth out this data that would otherwise be very oscillatory if values too quickly returned to zero, and does so in a meaningful way. Below we present the Hawkes process:
\begin{align}
    Hawkes(News(t)) = \sum_{i=0}^{\text{\# events before time t}} e^{-r * (t-t_e)}
\end{align}
where $i$ is the index of events (iterating from the first event to the most recent before time t), r is a decay constant calculated using the half-life being used, $t_e$ is the time of event currently being summed, and $t$ is time input to the Hawkes process. We let $\mathcal{T}^*$ and $\mathcal{N}^*$ denote the output of the Hawkes process over the time when receiving $\mathcal{T}$ and $\mathcal{N}$ as input, respectively.

Thus, for a given two week window representing row $i$ of the heatmap, that has a starting time $s_i$, ending time $e_i=s_i+w$, and offset $o_j \in \mathcal{O}$, we select the section of the Hawkes processed tweet sequence $\mathcal{T}^*(s_i, e_i)$ and news article sequence $\mathcal{N}^*(s_i+o_j,e_i+o_j)$ to discover the correlation $h_{ij}$ for the cell $(i,j)$ in the heatmap, which is defined as follows:
\begin{align}
    h_{ij} = correlation(\mathcal{T}^*(s_i, e_i),\mathcal{N}^*(s_i+o_j,e_i+o_j))
\end{align}
which can be visualized in Figure~\ref{fig:heatmapExplanation} where the news articles are shifted according to offset $o_j$ when aligned with the tweets for row $i$. Note that the entire heatmap is filled in using the same process described above by varying $i$ and $j$.

\begin{figure}[t]
 \centering
  \hspace*{-0.5cm}\includegraphics[width=0.53\textwidth]{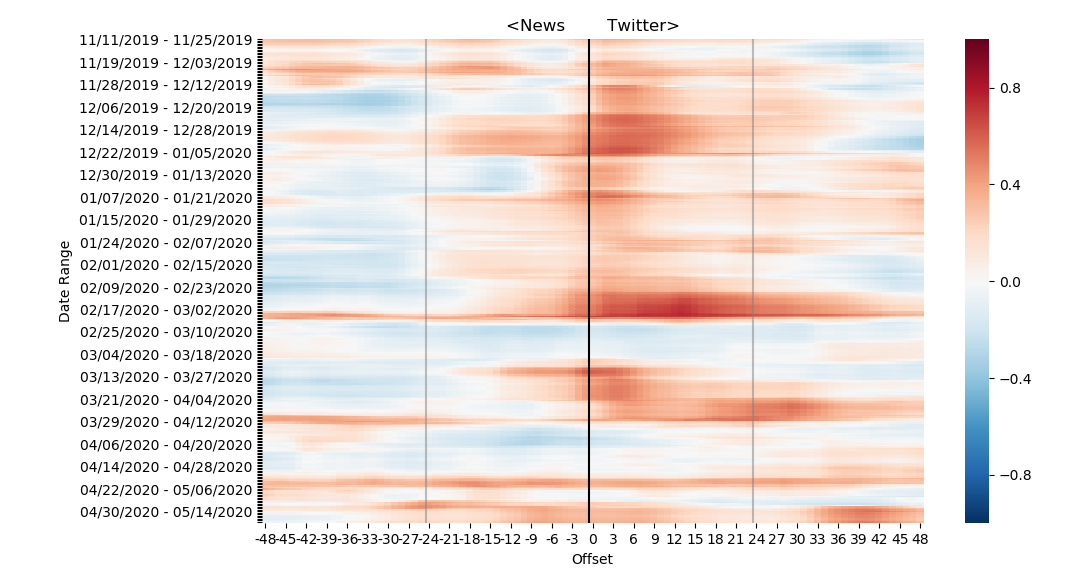}
  \vskip -2ex
 \caption{Andrew Yang windowed correlation heatmap.}
 \label{fig:AndrewYangHeatmap}
   \vskip -1ex
 \end{figure}
 
 \begin{figure}[t]
 \centering
  \hspace*{-0.5cm}\includegraphics[width=0.53\textwidth]{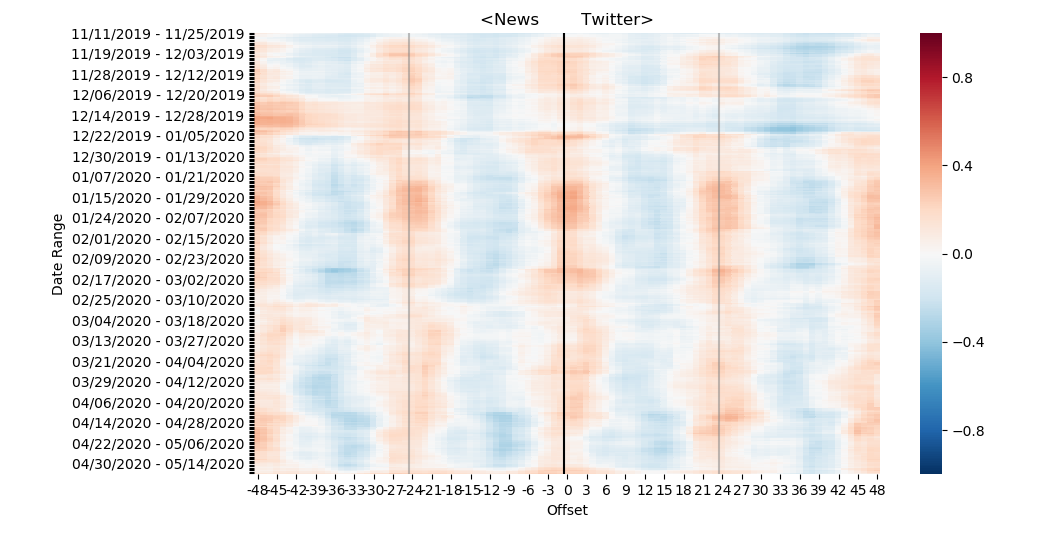}
    \vskip -2ex
 \caption{Donald Trump windowed correlation heatmap.}
 \label{fig:DonaldTrumpHeatmap}
   \vskip -1ex
 \end{figure}
 
  \begin{figure}[t]
 \centering
 \vskip -2ex
  \hspace*{-0.5cm}\includegraphics[width=0.53\textwidth]{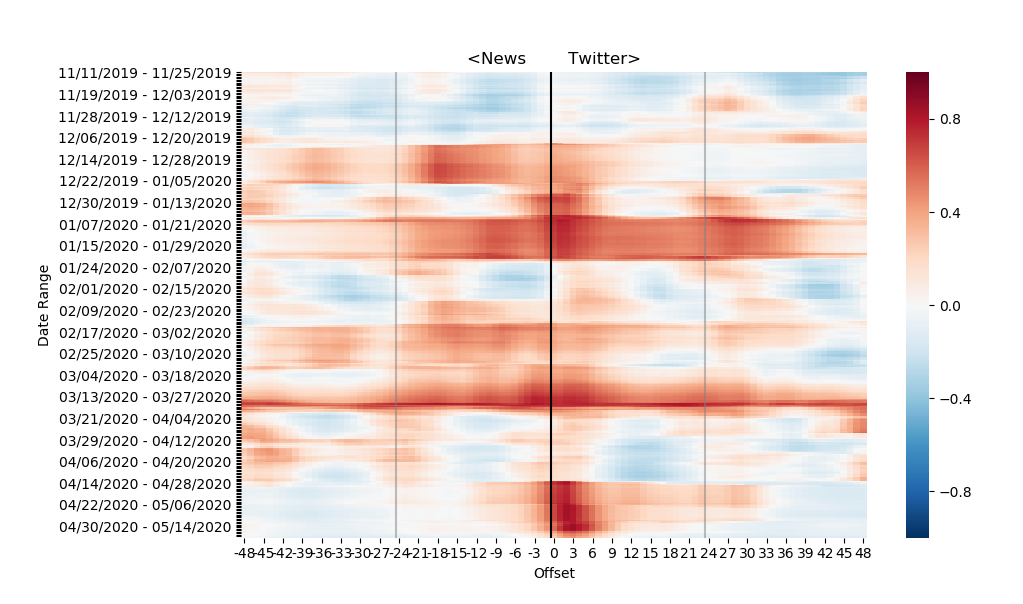}
    \vskip -2ex
 \caption{Elizabeth Warren windowed correlation heatmap.}
 \label{fig:ElizabethWarrenHeatmap}
   \vskip -1ex
 \end{figure}

Here we present the visualized heatmaps in Figures \ref{fig:AndrewYangHeatmap}, \ref{fig:DonaldTrumpHeatmap}, and \ref{fig:ElizabethWarrenHeatmap}, for Andrew Yang, Donald Trump, and Elizabeth Warren. Note that to the left side of the figures, the news is leading Twitter and on the right side Twitter is leading the news. Looking at the figures from top to bottom follows the passage of time. Red indicates areas of higher correlation, white is no correlation, and blue is a negative correlation. These figures show over time how Twitter and news datasets relate with each other by visualizing how the correlation between them changes over time and with different lags.

 Andrew Yang's heatmap is shown in Figure \ref{fig:AndrewYangHeatmap}. It can be seen that there are high correlation sections, and also sections that do not have a strong correlation no matter what the lag. Overall, there appears to be stronger correlations on the right side of the figure, indicating that on average, Andrew Yang's twitter base has a stronger influence on the news than the other way around. The darkest red spot on the heatmap is in mid-February when Andrew Yang dropped out of the presidential race. Interestingly, this high area of correlation is on the Twitter-leading side of the figure, indicating that there was a large response to his announcement on Twitter before there was a large news coverage response.
 
 Donald Trump's heatmap is shown in Figure \ref{fig:DonaldTrumpHeatmap}. There are segments of higher correlation with no shift, and also at exact increments of one day shifted. Donald Trump also has the most stable document publishing rates both on Twitter and news. This heatmap suggests that there is a daily schedule of more popular times and less popular times that is repeated consistently and daily. This is why any shift seems to quickly decrease the correlation, and the peaks of correlation are all at daily increments. It does not matter how far the data is shifted, as long as it is a period of 24 hours the daily schedule will line up, and the correlation will be at its maximum. This is supported by the fact that Trump's Tweet frequency is unusually high compared to other presidents and has a tendency to dominate the news cycle.\footnote{https://www.washingtonpost.com/politics/2020/05/12/how-much-trumps-presidency-has-he-spent-tweeting/}

Elizabeth Warren's heatmap is demonstrated in Figure \ref{fig:ElizabethWarrenHeatmap}. We are not able to observe a clear trend as seen with Andrew Yang (i.e., Figure \ref{fig:AndrewYangHeatmap}), but it does seem to be leaning slightly in the news influence Twitter direction. There are a lot of instances where the high correlation periods stretch far in both directions, indicating a period where the candidate is experiencing an unusually steady period of coverage. However, Warren has a few patches that exist only on the news leading side of the figure, indicating that overall news coverage seems to have a stronger influence on Twitter information than the other way around for Elizabeth Warren. We note the dark red spot in April started around the time when she endorsed Joe Biden as the Democratic nominee.

\section{Granger Causality}
While the previous section analyzes areas of influence, and how these areas change over time, it does not answer the more general question ``Does News impact social media more, or vice versa?'' The fine granularity of the previous analysis was helpful, but we perform a more general and grounded analysis.

\begin{figure}[t]
 \centering
  \includegraphics[width=0.46\textwidth]{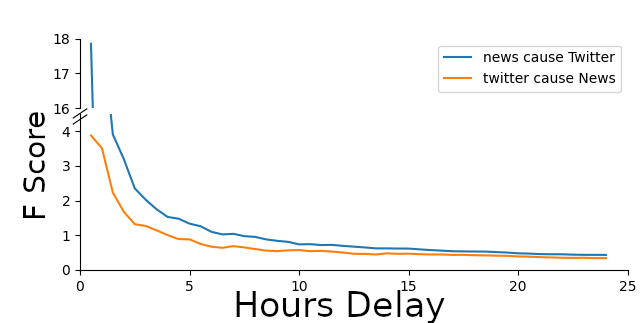}
  \vskip -1ex
 \caption{Granger Causality results.}
   \vskip -1ex
 \label{fig:Granger}
 \end{figure}

More specifically, we harness the Granger Causality test~\cite{granger1969investigating} as a way to show potential causalities between two sets of time series data, which has been heavily used for analyzing temporal data/events on Twitter~\cite{rizoiu2017tutorial}. The test is to see whether or not one time series (along with lags of that time series) are helpful in predicting the other time series. The Granger Causality is being applied to news and Twitter time series to explore the relationship between them. To increase granularity, the Hawkes process with a half-life equal to the average document period was ran on each series. In other words, to establish a principled half-life value, if on average news articles were appearing every 1 hour, then this was the value used. 
The test was run with lags up to 24 hours for both News cause Twitter and Twitter cause News. In Figure~\ref{fig:Granger} we show the resulting sum of squared residuals (SSR) based F-test results used in Granger Causality changes over the different time lags. We note that the average p-value for News influencing Twitter was $1.64 * 10^{-24}$, and the average p-value for Twitter influencing News was $1.53 * 10^{-12}$. Thus, although we have justification that Twitter does Granger-cause News, and News does Granger-cause Twitter, we observe a significantly smaller p-value for News influencing Twitter, as well as a significantly higher F score. This suggests that mainstream news has a larger impact on Twitter conversation than the other way around. While they both influence each other, it is clear that overall, Twitter influences the news less. One future direction to continue this work could be to explore if the news always interacts with Twitter in this way, or if that depends on the specific topic being discussed or difference per candidate.

 \begin{figure*}[t]
 \centering
  \includegraphics[width=.9\textwidth]{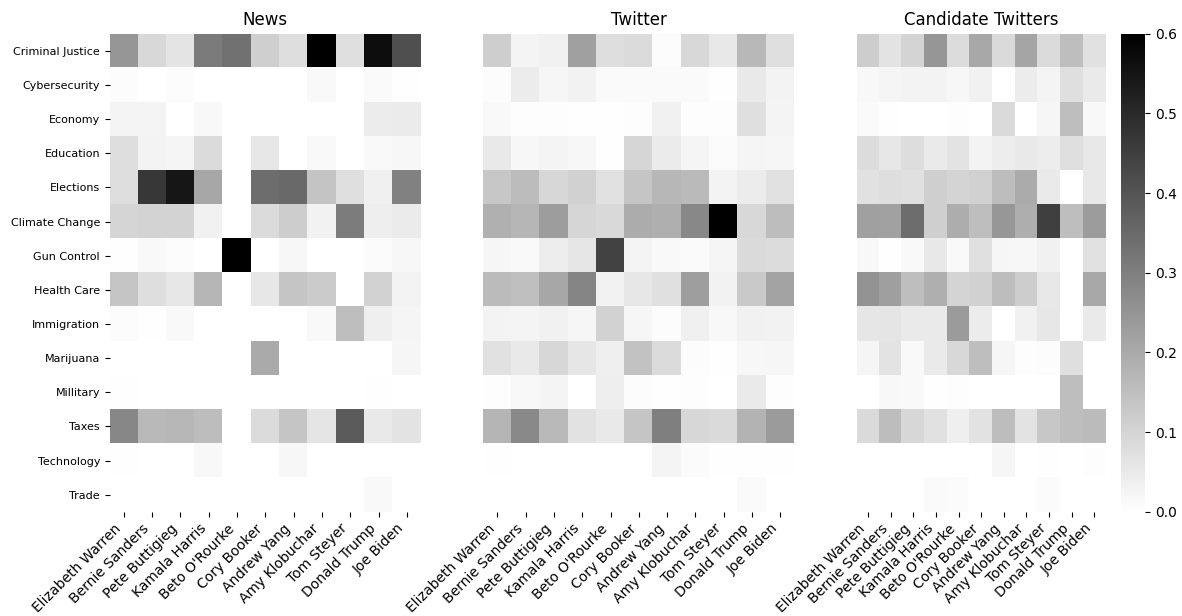}
    \vskip -1ex
 \caption{Topic distributions for each of the candidates across three data sources.}
 \label{fig:TopicVectors}
 \end{figure*}
 
We also note that Figure~\ref{fig:Granger} suggests that the influence of each of these news sources decreases rapidly.  After five hours, it is just a fraction of what it was originally. While obviously news articles and Tweets are still read for a long time after being published, it is clear in the two figures that any initial spike that one media source causes in the activity of the other usually happens within the first few hours.

\section{Topic MisMatch}

As a campaigning strategy, usually U.S. presidential candidates concentrate their rhetoric around certain topics, e.g., Donald Trump on immigration~\cite{reny2019vote} or Bernie Sanders on income and wealth inequality~\cite{CNNBernie}. However, this does not mean that a particular media perceives a candidate around the same topics. In other words, it is possible that a candidate is associated with some topics different from those he/she hinge upon. This creates a mismatch perception of what is important to a candidate and how he/she is portrayed on a media. In this part, we attempt to quantify this topic mismatch for each candidate across three sources, namely Twitter, mainstream media news, and candidates' own Tweets. The procedure is described in the following. 

First, we identify the topics associated with each candidate. To do this, we use the list of topics curated by \textit{politico.com}\footnote{https://www.politico.com/2020-election/candidates-views-on-the-issues/}. Then, we convert each topic to a numerical representation using the word2vec model~\cite{mikolov2013distributed}. Similarly, each text in all three sources (i.e., Twitter, mainstream news, and candidates' own Tweets) is converted to a numerical vector by taking the average of tokens' word2vec vectors in the text. The next step includes matching each text to a topic. To do this, each text is matched to the closest relating topic whose vector has the highest cosine similarity to the text's vector. To reduce the noise, we consider a minimum cutoff threshold, and any text that did not meet this threshold in cosine similarity to any of the topics was discarded. The distribution of topics for general Twitter, the candidates' Tweets, and mainstream news are shown in Figures~\ref{fig:TopicVectors}.

Finally, we normalize each candidates' data by calculating what percentage of the candidates' documents were in each category. This was done separately on each candidate to accommodate for the fact that some candidates had more documents than others. The distribution of topics for both general Twitter, the candidates' Twitter accounts, and News are shown in Figures~\ref{fig:TopicVectors}. In this figure, the darkness of each square represents the percentage of documents for the corresponding candidate that belongs to the corresponding topic. We make the below observations based on this figure.

First, we can observe some mismatches between the topics of interest for candidates themselves with those expressed on mainstream news and/or Twitter. Notably, for instance, gun control, while a topic of interest for Beto O'Rourke, is not a topic he tweets much about. This can be confirmed by his low value (bright color) for gun control in ``Candidate Twitters'', while mainstream media (i.e., ``News'') and ``Twitter'' have highly associated him with gun control. Next, for some topics, we see the consistency between different sources for all candidates. For instance, Cybersecurity has not been expressed much by almost all candidates. Likewise, mainstream media and Twitter have hardly associated it with any candidate. We believe this is related to the nature of the topic as some topics are essentially less ``uncontroversial" and consequently less mismatch occurs for them.

Our last observation is that while comparing News and Twitter matching distribution in the first and second panel of Figure~\ref{fig:TopicVectors}, respectively, in general, mainstream news shows more contrast. We attribute this more focused nature news articles where a particular topic about a candidate is discussed in more depth compared to the informal and occasionally erratic Twitter posts.

\section{Toxicity Analysis}

\begin{figure}[t]
 \centering
  \includegraphics[width=.48\textwidth]{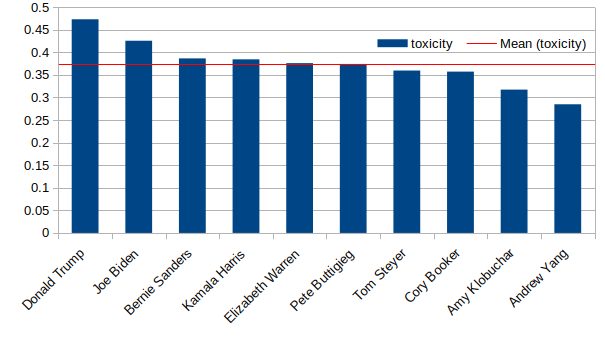}
  \vskip -1.5ex
 \caption{Toxicity of Tweets about Candidates.} 
 \label{fig:Toxicity}
 \vskip -1ex
 \end{figure}

Unfortunately, many online social media users experience some kind of toxic behavior such as sexism, racism, cyberbullying, etc~\cite{chatzakou2017mean}.  Toxicity escalates disagreements, irritates all parties involved, and hinders progress in a debate or conversation. Toxicity is especially prevalent in political conversations where people usually take a firm stand and are polarized~\cite{gruzd2014investigating}. Hence, to further deepen our understanding of the relationship between online social media and the election process, it is worthwhile to determine the toxicity level of tweets wherein candidates are being mentioned. To perform this analysis, we extract roughly 400,000 tweets per candidate that mentioned them and run this set through Google's Perspective API\footnote{https://support.perspectiveapi.com/} to get an average toxicity score between 0 and 1 for each candidate. The toxicity levels extracted per candidate are reported in Figure \ref{fig:Toxicity}.  Interestingly, the toxicity level closely follows the popularity of each candidate. That is, the more popular (well-known) a candidate is, the more toxic the conversion entailing that candidate is. One possible explanation for this is that popular candidates such as Donald Trump or Joe Biden are politically influential, their political statements naturally attract more people to the discussion, and consequently, the likelihood of toxic content is higher. This is further reinforced by the fact that the American political arena is currently extremely polarized especially when it comes to Donald Trump whose political stands and actions have largely widened the political polarization in the U.S.~\cite{abramowitz2019united}.

\section{Related Work}
Our work is related to streams of work related to investigating bias in media, social media's impact on elections, and agenda-setting. In fact, there are many works that have investigated media bias and how to detect it~\cite{automatedBiasDetection}. There has also been a lot of research on search bias, which is built on the idea that when searching online the results returned can be biased due to the data or the algorithms, including related to political science topics~\cite{searchBias}. In regards to social media's impact elections, works have shown that Twitter commonly has spikes of conversation about topics recently covered in the news regarding presidential elections~\cite{2012TwitterAnalysis} and have even attempted to predict election outcomes based on social media~\cite{tumasjan2010predicting}. The final related are of work is that of agenda-setting~\cite{parenti1993inventing}, where there have been recent works focused on this topic related to today's big data, and specifically online social media~\cite{vargo2018agenda,feezell2018agenda,lau2020media}.

\section{Conclusion}
Throughout this paper, we have demonstrated several different ways to interpret and display the data that we have been collecting from several media sources. By exploring the differences between liberal, central, and conservative news sources, the influence that social media and the news have on each other, topic mismatch, and toxicity, we have a better understanding of how the different media sources interact with each other and with their viewers when it comes to discussing the presidential candidates. Analyzing this data has shown some clear trends. For instance, one observation we made during sentiment analysis was that more central sources had lower overall sentiment than both liberal and conservative sources. We presented a few ideas about why this could be, including that liberal and conservative sources may have a tendency to write about topics that are favorable in their domain, which would lead to a larger overall sentiment. However, this is a proposed explanation for the observations made from the sentiment analysis, and aren't tested. For future work, it would be interesting to further test some of these proposed explanations and expand this project. 

\balance
\bibliographystyle{aaai}
\bibliography{references}

\begin{thebibliography}{}

\bibitem[\protect\citeauthoryear{Abramowitz and
  McCoy}{2019}]{abramowitz2019united}
Abramowitz, A., and McCoy, J.
\newblock 2019.
\newblock United states: Racial resentment, negative partisanship, and
  polarization in trump’s america.
\newblock {\em The ANNALS of the American Academy of Political and Social
  Science} 681(1):137--156.

\bibitem[\protect\citeauthoryear{Berkowitz}{}]{berkowitz1992sets}
Berkowitz, D.
\newblock Who sets the media agenda? the ability of policymakers to determine
  news decisions.
\newblock {\em Public opinion, the press, and public policy} 2:81--102.

\bibitem[\protect\citeauthoryear{Berkowitz}{1987}]{berkowitz1987tv}
Berkowitz, D.
\newblock 1987.
\newblock Tv news sources and news channels: A study in agenda-building.
\newblock {\em Journalism Quarterly} 64(2-3):508--513.

\bibitem[\protect\citeauthoryear{Bernhardt, Krasa, and
  Polborn}{2008}]{polarizationBiasModel}
Bernhardt, D.; Krasa, S.; and Polborn, M.
\newblock 2008.
\newblock Political polarization and the electoral effects of media bias.
\newblock {\em Journal of Public Economics} 92(5-6):1092--1104.

\bibitem[\protect\citeauthoryear{Chatzakou \bgroup et al\mbox.\egroup
  }{2017}]{chatzakou2017mean}
Chatzakou, D.; Kourtellis, N.; Blackburn, J.; De~Cristofaro, E.; Stringhini,
  G.; and Vakali, A.
\newblock 2017.
\newblock Mean birds: Detecting aggression and bullying on twitter.
\newblock In {\em Proceedings of the 2017 ACM on web science conference},
  13--22.

\bibitem[\protect\citeauthoryear{Cogburn and
  Espinoza-Vasquez}{2011}]{cogburn2011networked}
Cogburn, D.~L., and Espinoza-Vasquez, F.~K.
\newblock 2011.
\newblock From networked nominee to networked nation: Examining the impact of
  web 2.0 and social media on political participation and civic engagement in
  the 2008 obama campaign.
\newblock {\em Journal of political marketing} 10(1-2):189--213.

\bibitem[\protect\citeauthoryear{Dahlberg}{2007}]{dahlberg2007internet}
Dahlberg, L.
\newblock 2007.
\newblock The internet, deliberative democracy, and power: Radicalizing the
  public sphere.
\newblock {\em International Journal of Media \& Cultural Politics}
  3(1):47--64.

\bibitem[\protect\citeauthoryear{D’Avanzo, Pilato, and
  Lytras}{2017}]{trendTwitterRealOpinion}
D’Avanzo, E.; Pilato, G.; and Lytras, M.
\newblock 2017.
\newblock Using twitter sentiment and emotions analysis of google trends for
  decisions making.
\newblock {\em Program}.

\bibitem[\protect\citeauthoryear{Egan}{2015}]{CNNBernie}
Egan, M.
\newblock 2015.
\newblock {\em The Bernie Sanders plan to fix income inequality}.
\newblock Accessed: 2020-09-01.

\bibitem[\protect\citeauthoryear{Feezell}{2018}]{feezell2018agenda}
Feezell, J.~T.
\newblock 2018.
\newblock Agenda setting through social media: The importance of incidental
  news exposure and social filtering in the digital era.
\newblock {\em Political Research Quarterly} 71(2):482--494.

\bibitem[\protect\citeauthoryear{Flaxman, Goel, and
  Rao}{2016}]{flaxman2016filter}
Flaxman, S.; Goel, S.; and Rao, J.~M.
\newblock 2016.
\newblock Filter bubbles, echo chambers, and online news consumption.
\newblock {\em Public opinion quarterly} 80(S1):298--320.

\bibitem[\protect\citeauthoryear{Granger}{1969}]{granger1969investigating}
Granger, C.~W.
\newblock 1969.
\newblock Investigating causal relations by econometric models and
  cross-spectral methods.
\newblock {\em Econometrica: journal of the Econometric Society}  424--438.

\bibitem[\protect\citeauthoryear{Gruzd and Roy}{2014}]{gruzd2014investigating}
Gruzd, A., and Roy, J.
\newblock 2014.
\newblock Investigating political polarization on twitter: A canadian
  perspective.
\newblock {\em Policy \& internet} 6(1):28--45.

\bibitem[\protect\citeauthoryear{Hamborg, Donnay, and
  Gipp}{2019}]{automatedBiasDetection}
Hamborg, F.; Donnay, K.; and Gipp, B.
\newblock 2019.
\newblock Automated identification of media bias in news articles: an
  interdisciplinary literature review.
\newblock {\em International Journal on Digital Libraries} 20(4):391--415.

\bibitem[\protect\citeauthoryear{Hutto and Gilbert}{2015}]{vader}
Hutto, C., and Gilbert, E.
\newblock 2015.
\newblock Vader: A parsimonious rule-based model for sentiment analysis of
  social media text.

\bibitem[\protect\citeauthoryear{Jang \bgroup et al\mbox.\egroup
  }{2019}]{jang2019social}
Jang, S.~M.; Mckeever, B.~W.; Mckeever, R.; and Kim, J.~K.
\newblock 2019.
\newblock From social media to mainstream news: The information flow of the
  vaccine-autism controversy in the us, canada, and the uk.
\newblock {\em Health communication} 34(1):110--117.

\bibitem[\protect\citeauthoryear{Jungherr}{2014}]{jungherr2014twitter}
Jungherr, A.
\newblock 2014.
\newblock Twitter in politics: a comprehensive literature review.
\newblock {\em Available at SSRN 2865150}.

\bibitem[\protect\citeauthoryear{Kulshrestha \bgroup et al\mbox.\egroup
  }{2017}]{searchBias}
Kulshrestha, J.; Eslami, M.; Messias, J.; Zafar, M.~B.; Ghosh, S.; Gummadi,
  K.~P.; and Karahalios, K.
\newblock 2017.
\newblock Quantifying search bias: Investigating sources of bias for political
  searches in social media.
\newblock In {\em Proceedings of the 2017 ACM Conference on Computer Supported
  Cooperative Work and Social Computing},  417--432.

\bibitem[\protect\citeauthoryear{Kushin and Yamamoto}{2010}]{kushin2010did}
Kushin, M.~J., and Yamamoto, M.
\newblock 2010.
\newblock Did social media really matter? college students' use of online media
  and political decision making in the 2008 election.
\newblock {\em Mass Communication and Society} 13(5):608--630.

\bibitem[\protect\citeauthoryear{Lau, Rogers, and Love}{2020}]{lau2020media}
Lau, R.~R.; Rogers, K.; and Love, J.
\newblock 2020.
\newblock Media effects in the viewer’s choice era: Testing revised
  agenda-setting and priming hypotheses.
\newblock {\em Political Communication}  1--23.

\bibitem[\protect\citeauthoryear{Liu and Zhang}{2012}]{liu2012survey}
Liu, B., and Zhang, L.
\newblock 2012.
\newblock A survey of opinion mining and sentiment analysis.
\newblock In {\em Mining text data}. Springer.
\newblock  415--463.

\bibitem[\protect\citeauthoryear{Metzgar and Maruggi}{2009}]{metzgar2009social}
Metzgar, E., and Maruggi, A.
\newblock 2009.
\newblock Social media and the 2008 us presidential election.
\newblock {\em Journal of New Communications Research} 4(1).

\bibitem[\protect\citeauthoryear{Mikolov \bgroup et al\mbox.\egroup
  }{2013}]{mikolov2013distributed}
Mikolov, T.; Sutskever, I.; Chen, K.; Corrado, G.~S.; and Dean, J.
\newblock 2013.
\newblock Distributed representations of words and phrases and their
  compositionality.
\newblock In {\em Advances in neural information processing systems},
  3111--3119.

\bibitem[\protect\citeauthoryear{Nadeau \bgroup et al\mbox.\egroup
  }{2008}]{nadeau2008election}
Nadeau, R.; Nevitte, N.; Gidengil, E.; and Blais, A.
\newblock 2008.
\newblock Election campaigns as information campaigns: who learns what and does
  it matter?
\newblock {\em Political Communication} 25(3):229--248.

\bibitem[\protect\citeauthoryear{Newman}{2011}]{newman2011mainstream}
Newman, N.
\newblock 2011.
\newblock Mainstream media and the distribution of news in the age of social
  media.

\bibitem[\protect\citeauthoryear{Parenti}{1993}]{parenti1993inventing}
Parenti, M.
\newblock 1993.
\newblock {\em Inventing reality: The politics of news media}.
\newblock St. Martin's Press New York.

\bibitem[\protect\citeauthoryear{Reny, Collingwood, and
  Valenzuela}{2019}]{reny2019vote}
Reny, T.~T.; Collingwood, L.; and Valenzuela, A.~A.
\newblock 2019.
\newblock Vote switching in the 2016 election: How racial and immigration
  attitudes, not economics, explain shifts in white voting.
\newblock {\em Public Opinion Quarterly} 83(1):91--113.

\bibitem[\protect\citeauthoryear{Rizoiu \bgroup et al\mbox.\egroup
  }{2017}]{rizoiu2017tutorial}
Rizoiu, M.-A.; Lee, Y.; Mishra, S.; and Xie, L.
\newblock 2017.
\newblock A tutorial on hawkes processes for events in social media.
\newblock {\em arXiv preprint arXiv:1708.06401}.

\bibitem[\protect\citeauthoryear{Roberts and McCombs}{1994}]{roberts1994agenda}
Roberts, M., and McCombs, M.
\newblock 1994.
\newblock Agenda setting and political advertising: Origins of the news agenda.
\newblock {\em Political communication} 11(3):249--262.

\bibitem[\protect\citeauthoryear{Tumasjan \bgroup et al\mbox.\egroup
  }{2010}]{tumasjan2010predicting}
Tumasjan, A.; Sprenger, T.~O.; Sandner, P.~G.; and Welpe, I.~M.
\newblock 2010.
\newblock Predicting elections with twitter: What 140 characters reveal about
  political sentiment.
\newblock In {\em Fourth international AAAI conference on weblogs and social
  media}.
\newblock Citeseer.

\bibitem[\protect\citeauthoryear{Vargo, Guo, and
  Amazeen}{2018}]{vargo2018agenda}
Vargo, C.~J.; Guo, L.; and Amazeen, M.~A.
\newblock 2018.
\newblock The agenda-setting power of fake news: A big data analysis of the
  online media landscape from 2014 to 2016.
\newblock {\em New media \& society} 20(5):2028--2049.

\bibitem[\protect\citeauthoryear{Vosoughi, Roy, and
  Aral}{2018}]{vosoughi2018spread}
Vosoughi, S.; Roy, D.; and Aral, S.
\newblock 2018.
\newblock The spread of true and false news online.
\newblock {\em Science} 359(6380):1146--1151.

\bibitem[\protect\citeauthoryear{Wang \bgroup et al\mbox.\egroup
  }{2012}]{2012TwitterAnalysis}
Wang, H.; Can, D.; Kazemzadeh, A.; Bar, F.; and Narayanan, S.
\newblock 2012.
\newblock A system for real-time twitter sentiment analysis of 2012 u.s.
  presidential election cycle.
\newblock In {\em Proceedings of the ACL 2012 System Demonstrations}, ACL '12,
  115–120.
\newblock USA: Association for Computational Linguistics.

\bibitem[\protect\citeauthoryear{Weare}{2002}]{weare2002internet}
Weare, C.
\newblock 2002.
\newblock The internet and democracy: The causal links between technology and
  politics.
\newblock {\em International Journal of Public Administration} 25(5):659--691.

\end{thebibliography}

\end{document}